\documentclass[aps,prl,twocolumn,superscriptaddress,amsmath,amssymb]{revtex4}
\usepackage{graphicx}
\usepackage{amsmath}
\usepackage{pstricks}
\usepackage{tikz}
\usepackage{float}

\begin{document}

\title{Emergent Gauge Field for a Chiral Bound State on Curved Surface}
\author{Zhe-Yu Shi}
\affiliation{Institute for Advanced Study, Tsinghua University, Beijing, 100084, China}
\author{Hui Zhai}
\affiliation{Institute for Advanced Study, Tsinghua University, Beijing, 100084, China}
\date{\today}
\begin{abstract}
In this letter we show that there emerges a gauge field for two attractive particles moving on a curved surface when they form a chiral bound state. By solving a two-body problem on a sphere, we show explicitly that the center-of-mass wave functions of such deeply bound states are monopole harmonics instead of spherical harmonics. This indicates that the bound state experiences a gauge field identical to a magnetic monopole at the center of the sphere, with the monopole charge equal to the quantized relative angular momentum of this bound state. We show that this emergent gauge field is due to the coupling between the center-of-mass and the relative motion on curved surfaces. Our results can be generalized to an arbitrary curved surface where the emergent magnetic field is exactly the local Gaussian curvature. This result establishes an intriguing connection between space curvature and gauge field, paves an alternative way to engineer topological state with space curvature, and may be observed in cold atom system.
\end{abstract}
\maketitle

Emergent physics is one of the most important concept in the modern physics, and one of the most intriguing example is the emergent gauge field in the low-energy physics. That is to say, the original model does not contain a gauge field explicitly, but a gauge theory appears in the low-energy effective theory when tracing out some high-energy degree of freedoms. Emergent gauge field has appeared, for instance, in the low-energy field theory description of strongly correlated materials such as doped Mott insulators and spin liquids \cite{Wen} or Dirac fermions in graphene under local distortions \cite{graphene}. Here we propose a new scenario that when two particles form a chiral bound state on a two-dimensional curved surface, the center-of-mass motion of this pair, as the remaining low-energy degree of freedom, experiences a gauge field originated from the space curvature. This result brings about interesting connection between space curvature and gauge field in a simple system.

We consider a system consisting two particles interacting via a short-range potential. Such a problem can be treated by partial wave expansion, and generally the $s$-wave channel is the most dominating one for low-energy collisions. Nevertheless, when the potential is tuned to a $p$-wave or other high partial wave resonances, the interaction in this high partial wave channel can be very strong and a chiral bound state with non-zero relative angular momentum can form. This kind of two-body problem has been extensively studied in various circumstances in cold atom systems, such as in reduced dimensions \cite{CIR1,CIR2}, or in the presence of gauge field or spin-orbit coupling \cite{so_coupling}. It has been found that both the dimensionality and the gauge field can strongly affect the low-energy behavior of scattering phase shift and bound states. However, such a problem has not been considered in curved spaces and the effect of space curvature on bound state behavior has not been addressed before. This will be the main issue to address in this work. We should also stress that our results focus on the strong pairing regime, which is complementary to previous discussions of BCS superfluids \cite{BCS1,BCS2} or superfluid vortex on curved surfaces.\cite{vortex}

\begin{figure}[tbp]\centering
\includegraphics[width=3.4in]
{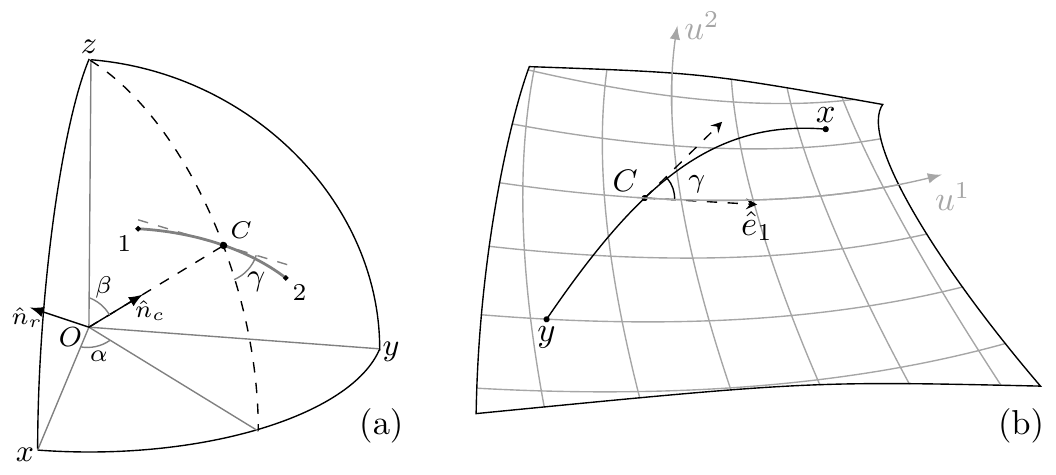}
\caption{(a): A schematic plot of coordinates $(\alpha,\beta,\gamma,\theta)$ for two particles on a sphere. Angle $\theta$(which is not shown on the plot) is the angle between $\mathbf{r}_1$ and $\mathbf{r}_2$ (b): A schematic plot of coordinates $(q^1,q^2,\gamma,r)$ for two particles on a general curved surface. Here $q_1$ and $q_2$ correspond to $\alpha$ and $\beta$ in (a) and $r$ corresponds to $R\theta$ in (a).  }\label{fig1}
\end{figure}

{\bf Emergent Gauge Filed in Quantum-Mechanical Two-Body Problem on a Sphere.} Before considering a general situation of a curved surface, we illustrate the physics by solving a concrete model of two particles confined on a sphere (with radius $R$) with short-range interaction (with range denoted by $r_0$). Following the standard procedure of solving a two-body problem with short-range interaction, hereafter we first solve this non-interacting problem when the distance between two atoms are larger than $r_0$ and then fix the entire wave function by matching the short-range Bethe-Peierls boundary condition. In detail, it follows following steps:

\textit {1. Defining Coordinates.} The two-body system can be described by the coordinates of two particles ${\bf r}_1$ and ${\bf r}_2$. For the convenience of separating the center-of-mass and the relative coordinates, we introduce another set of coordinates $(\alpha,\beta,\gamma,\theta)$. We define unit vectors $\hat{n}_c$ and $\hat{n}_r$ as $\hat{n}_c=\frac{\mathbf{r}_1+\mathbf{r}_2}{|\mathbf{r}_1+\mathbf{r}_2|}=\frac{\mathbf{r}_1+\mathbf{r}_2}{2R\cos{\theta/2}}$
and $\hat{n}_r=\frac{\mathbf{r}_1-\mathbf{r}_2}{|\mathbf{r}_1-\mathbf{r}_2|}=\frac{\mathbf{r}_1-\mathbf{r}_2}{2R\sin{\theta/2}}$ , which represent the directions of the center of mass and the relative positions, respectively, and $\theta$ is the angle between the positions of two particles $\mathbf{r}_1$ and $\mathbf{r}_2$. Since $\hat{n}_c$ and $\hat{n}_r$ are orthogonal to each other, we can set a body-fixed frame $Ox'y'z'$ where $\hat{n}_c$ is the $z'$ axis, $\hat{n}_r$ is the $y'$ axis and $x'$ axis is determined by the right hand rule. The body-fixed frame $Ox'y'z'$ is related to the space-fixed frame $Oxyz$ by a rotation matrix $R$, i.e.
$(\hat{e}_{x'},\hat{e}_{y'},\hat{e}_{z'})^T=R(\hat{e}_{x},\hat{e}_{y},\hat{e}_{z})^T$,
and the rotation matrix $R$ can be parameterized by three Euler angles $(\alpha,\beta,\gamma)$ \cite{Euler angles}.

Fig.\ref{fig1}(a) shows the geometric meaning of $(\alpha,\beta,\gamma,\theta)$. We define the center-of-mass point $\mathcal{C}$ as the middle point of the geodesic (the great circle), and $\alpha$ and $\beta$ are the azimuthal and polar angles of point $\mathcal{C}$. $\gamma$ is the rotational angle of the geodesic along ${\bf n}_c$, it specifies the orientation of the geodesic. The angle $\theta$ represents the relative distance between the two atoms, and is proportional to the length of the geodesic. That is to say, the coordinates ($\alpha,\beta$) describe the center of mass position, while the coordinates $(\gamma,\theta)$ describe the relative motion.

\textit{2. Expressing the Kinetic Energy Operator in the New Coordinates.}
The two-body system $(\mathbf{r}_1,\mathbf{r}_2)$ can be regarded as a single point moving on the product manifold $S^2\times S^2$ equipped with metric $ds^2=d\mathbf{r}_1^2+d\mathbf{r}_2^2$. Under the coordinates $(\alpha,\beta,\gamma,\theta)$, the metric tensor becomes $ds^2=d\mathbf{r}_1^2+d\mathbf{r}_2^2=g_{ij}du^idu^j$,
where $(u^1,u^2,u^3,u^4)=(\alpha,\beta,\gamma,\theta)$, where $g_{ij}$ is listed in the appendix. Now the kinetic energy operator becomes (we set $\hbar=m=R=1$)
\begin{eqnarray}
\hat{T}=\frac{\mathbf{L}_1^2}{2}+\frac{\mathbf{L}_2^2}{2}=-\frac{1}{2}\frac{1}{\sqrt{g}}\partial_i(\sqrt{g}g^{ij}\partial_j)\label{laplace},
\end{eqnarray}
where $g=\sin^2\beta\sin^2\theta$ is the determinant of matrix $g_{ij}$, $g^{ij}$ is the matrix inverse of the $g_{ij}$. After some straightforward simplification, we obtain
\begin{align}
\hat{T}=\frac{J_1^2}{4}+\frac{J_2^2}{4\cos^2\frac{\theta}{2}}+\frac{J_3^2}{4\sin^2\frac{\theta}{2}}-\frac{1}{\sin\theta}\frac{\partial}{\partial\theta}\bigg{(}\sin\theta\frac{\partial}{\partial\theta}\bigg{)}, \label{H_rot}
\end{align}
where $J_1,J_2,J_3$ are the angular momentums along three body-fixed axis $x'$,$y'$ and $z'$. It is worth mentioning that $J_i$ only depends on Euler angles $(\alpha,\beta,\gamma)$, and their explicit expressions are listed in the appendix.

In the limit $\theta\rightarrow 0$, $\hat{T}$ becomes
\begin{eqnarray}
\hat{T}\simeq\frac{J_1^2+J_2^2}{4}-\frac{1}{\theta}\frac{\partial}{\partial\theta}\bigg{(}\theta\frac{\partial}{\partial\theta}\bigg{)}
+\frac{J_3^2}{\theta^2}.
\end{eqnarray}
In this limit, the 2-body system is effectively living on the tangent plane of the sphere. By comparing $\hat{T}$ with the kinetic energy operator on a flat 2D plane,
we can see that $\theta$ corresponds to the relative distance between two particles, $J_3$ is the relative angular momentum $l$ (therefore, hereafter we should use $l$ to denote quantum number of $J_3$,) and $J_1$ and $J_2$ represent the center of mass momentum. This also suggests that we can still apply the short-range boundary condition for a flat 2D plane to our problem.


\textit{3. Eigen-Wavefunction of the Kinetic Energy.} To determine the eigenstates of the kinetic energy term, we expand the wave function in terms of Wigner D-matrices, which are often used to solve the spectrum of a rigid rotor \cite{Landau}. The Wigner D-matrix $D_{ml}^j(\alpha,\beta,\gamma)^*$ is a common eigenstate of operators $\mathbf{J}^2$, $J_z$ and $J_3$ with eigenvalues $j(j+1)$, $m$ and $l$, respectively. Here $J_z=-i\partial_\alpha$ represents the angular momentum along the space-fixed $z$ axis. $\hat{T}$ has a simple form under basis $D_{ml}^j$ because
\begin{eqnarray}
\frac{J_1^2}{4}+\frac{J_2^2}{4\cos^2\theta/2}+\frac{J_3^2}{4\sin^2\theta/2}=A\mathbf{J}^2+BJ_3^2-C(J_+^2+J_-^2),\nonumber
\end{eqnarray}
where $J_\pm=J_1\pm iJ_2$ are the lowering and raising operators of $J_3$. $A$, $B$ and $C$ are simple functions of $\theta$ (See appendix for definitions).
From the above equation, we can see that the first two terms are diagonal under basis $D^j_{ml}$, while $J_{\pm}^2$ can couple $l$ to $l\mp2$.

For instance, if we consider a $p$-wave interaction, $l$ should take the value of $\pm 1$ at short distance and can be coupled to $l=\pm 3, \cdots$ at long distance. Therefore the wave functoin can be written as
\begin{eqnarray}
\psi^{\pm}=\varphi_1^{\pm}(\theta)\frac{D^{j*}_{m,1}\pm D^{j*}_{m,-1}}{\sqrt{2}}+\varphi_3^{\pm}(\theta)\frac{D^{j*}_{m,3}\pm D^{j*}_{m,-3}}{\sqrt{2}}+\ldots.\nonumber\\\label{ansatz2}
\end{eqnarray}
Here the superscript $\pm$ stands for two sets of solutions with different parity. Here we note that the expansion only contains a finite number of terms since $l$ cannot exceed the total angular momentum $j$.


\textit {4. Matching the Short-Range Boundary Condition and Determining the Wave Function.}For $p$-wave interaction on a 2D plane, the boundary condition is
\begin{eqnarray}
\phi_{\text{2D}}(r)\sim\bigg{(}\frac{1}{r}-\frac{\pi r}{4s}\bigg{)}-\frac{E_{\text{rel}}}{2}\log\frac{r}{r_0},\label{boundary}
\end{eqnarray}
where $s$ is the scattering area, $r_0$ is the effective range for the $p$-wave scattering, and $E_{\text{rel}}$ is the energy for relative motion. $s$ can be tuned by a magnetic Feshbach resonance in an ultracold atom system. The eigenenergy $E$ can then be calculated by expanding the wavefunctions Eq.(\ref{ansatz2}) in the short-range limit then comparing them with the the boundary condition Eq.(\ref{boundary}).

\begin{figure}[tbp]\centering
\includegraphics[width=3.5in]
{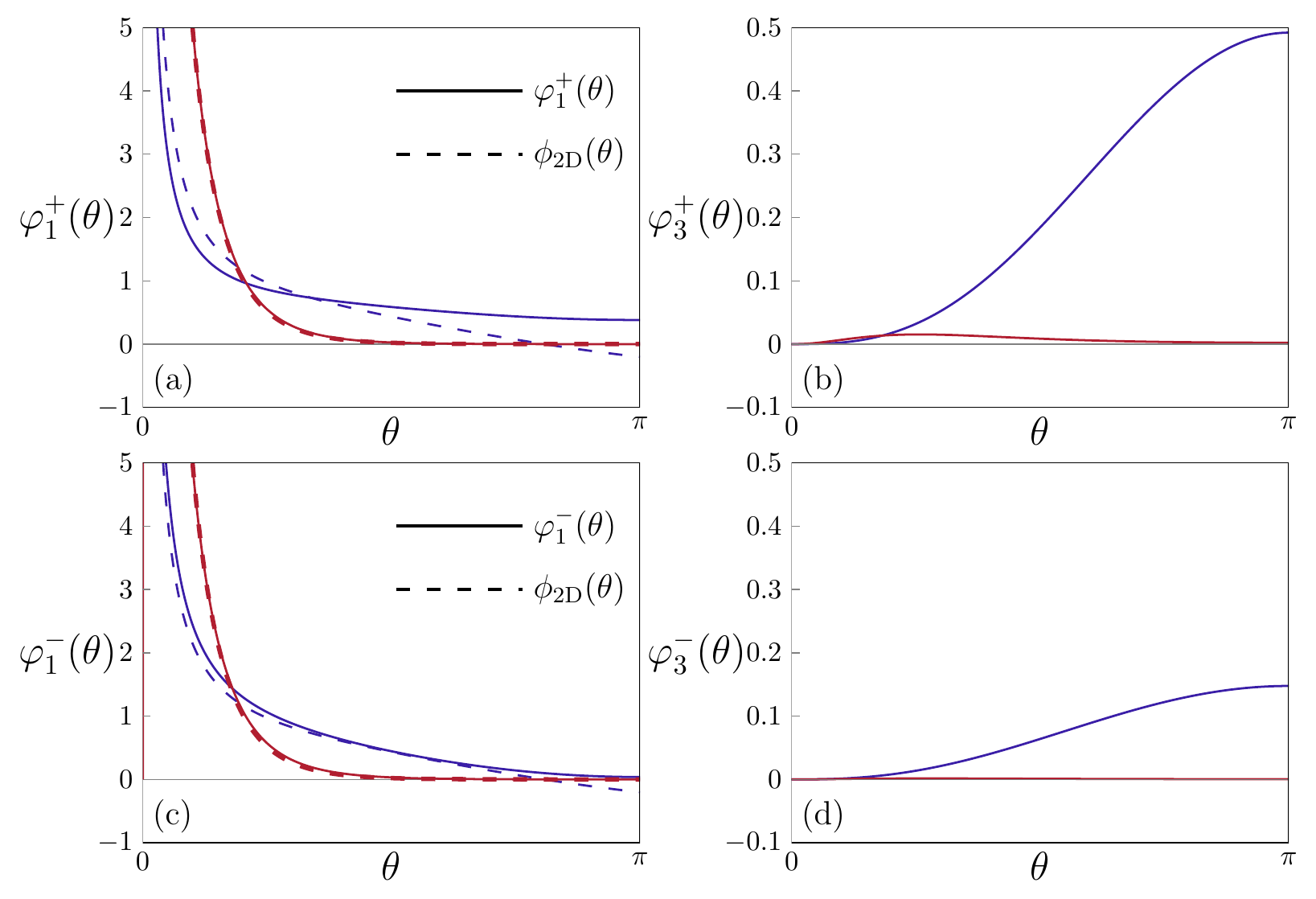}
\caption{The wave functions $\varphi_1^{\pm}$ ((a) and (c)) and $\varphi_3^{\pm}$ ((b) and (d)) for total angular momentum $j=3$. Solid blue lines are for a shallow bound state with $E=-3.5\hbar^2/(mR^2)$, while solid red lines are for a deeply bound state with $E=-10\hbar^2/(mR^2)$. The dashed line is the bound state wave function $\phi_{\text{2D}}(r)$ with $r=R\theta$ for 2D plane with same short-range boundary condition. This plot shows that for a deeply bound state, both $\varphi_{1}^{+}$ and $\varphi_{1}^{-}$ approach the same $\phi_\text{2D}$ and $\varphi_3^{\pm}$ vanish. }\label{fig2}
\end{figure}

In Fig.\ref{fig2}, we plot the wavefunctions for total angular momentum $j=3$.
As one can see, it turns out that, for a $p$-wave deeply bound state, both $\varphi_1^{\pm}$ approach the same wave function and $\varphi_{l>1}^\pm$ vanish as
\begin{eqnarray}
\varphi_l^\pm(\theta)\simeq\left\{
                         \begin{array}{ll}
                           \phi_{\text{2D}}(\theta), & l=1, \\
                           0, & l>1,
                         \end{array}
                       \right.
\end{eqnarray}
where $\phi_{\text{2D}}(r)$ is the radial wavefunction of the same system with same interaction on a 2D plane. Moreover $\psi^{\pm}$ becomes two nearly degenerate ground states. Thus, we can define two new eigenstates as $\psi_1=\frac{\psi^++\psi^-}{\sqrt{2}}$ and $\psi_{-1}=\frac{\psi^+-\psi^-}{\sqrt{2}}$, which become
\begin{align}
\psi_1&\simeq\phi_{\text{2D}}(\theta)e^{i\gamma}D^j_{m,1}(\alpha,\beta,0)^*,\nonumber\\
\psi_{-1}&\simeq\phi_{\text{2D}}(\theta)e^{-i\gamma}D^j_{m,-1}(\alpha,\beta,0)^*.
\end{align}
where we use $D^j_{m,\pm 1}(\alpha,\beta,\gamma)^*=e^{\pm i\gamma}D^j_{m, \pm 1}(\alpha,\beta,0)^*$. In fact, it can be shown that for a deeply bound state formed in any $l$-partial wave channel, the wave function is given by
\begin{eqnarray}
\psi_{\pm l}(\alpha,\beta,\gamma,\theta)\simeq D^j_{m,\pm l}(\alpha,\beta,0)^*\psi_{\text{2D}}(\gamma,\theta),
\end{eqnarray}

This is a central result of this part. We see that in the limit of a deeply bound state, the center-of-mass and the relative motion become effectively decoupled and the wavefunction can be written as a product of the relative and the center-of-mass wavefunctions. The center-of-mass wave function is the most intriguing and suggestive part. It is represented by a Wigner D-matrix. Mathematically, it is known that the Wigner D-matrices are related to the monopole harmonics $Y_{-l,j,m}$ by a gauge transformation introduced by Wu and Yang\cite{Wu and Yang, Wu and Yang2}. They are the eigenstates of a charged particle moving around a magnetic monopole. This suggests that there emerges a gauge field in our system which couples to the center-of-mass motion, and the charge of this monopole is the quantum number $-l$. That is to say, it exists for $p$-wave or other higher partial wave bound states but not for $s$-wave. And for bound states with opposite angular momentum $l$ or $-l$, they experience opposite monopole charge.


\textbf{Emergent Gauge Field in the Classical Hamiltonian.} The quantum mechanical calculation has shown ambiguously the emergence of a gauge field coupled to a chiral bound state. To gain a more intuitive understanding of this result, we now show, with a classical Hamiltonian of the same system, that this gauge field emerges from separating the center-of-mass and the relative degrees of freedom in the kinetic energy term. With this derivation, the geometric meaning of this gauge field also becomes clear and it can be generalized to arbitrary two-dimensional manifolds.

\textit{Gauge Field in the Classical Hamiltonian. } The classical kinetic energy is given by $T=\frac{1}{2}g_{ij}\dot{u}^i\dot{u}^j$, it can be separated into $T_{\text{c}}+T_{\text{rel}}$ as
\begin{eqnarray}
T_{\text{c}}&=\frac12\left(
           \begin{array}{cc}
             \dot{\alpha}, & \dot{\beta} \\
           \end{array}
         \right)h_{\mu\nu}\left(
                             \begin{array}{c}
                               \dot{\alpha} \\
                               \dot{\beta} \\
                             \end{array}
                           \right),\nonumber\\
T_{\text{rel}}&=\frac12I_3(\dot{\gamma}+\cos\beta\dot{\alpha})^2+\frac{1}{4}\dot{\theta}^2.
\end{eqnarray}
where $I_3=2\sin^2\frac{\theta}{2}$ is the moment of inertia along $z'$ axis, $h_{\mu\nu}$ is a 2 by 2 matrix whose elements are listed in the appendix. The angular velocity along $z'$-axis is $\omega_3=\dot{\gamma}+\cos\beta\dot{\alpha}=\dot{\gamma}+A_{\alpha}\dot{\alpha}$, where $A_\alpha=\cos\beta$. Introducing the conjugate momenta as $p_i=\partial T/\partial u^i$ as
\begin{align}
&\left(
  \begin{array}{c}
    p_\alpha \\
    p_\beta \\
  \end{array}
\right)=h_{\mu\nu}\left(
                    \begin{array}{c}
                      \dot{\alpha} \\
                      \dot{\beta} \\
                    \end{array}
                  \right)+I_3 A_\alpha\left(
                                        \begin{array}{c}
                                          \dot{\gamma}+A_\alpha\dot{\alpha} \\
                                          0 \\
                                        \end{array}
                                      \right),\nonumber\\
&L\mathrel{\mathop:}=p_\gamma=I_3(\dot{\gamma}+A_\alpha\dot{\alpha}),\qquad p_\theta=\frac12\dot{\theta},
\end{align}
the classical Hamiltonian can be rewritten as
\begin{eqnarray}
H=\frac12(
             p_\alpha-L A_\alpha,\  p_\beta)
            h^{\mu\nu}\left(
                            \begin{array}{c}
                              p_\alpha-L A_\alpha \\
                              p_\beta \\
                            \end{array}
                          \right)
\frac{L^2}{2I_3}+p_\theta^2+\dots,\nonumber\label{classical_H}
\end{eqnarray}
where $h^{\mu\nu}$ is the inverse of $h_{\mu\nu}$, and $\dots$ represents the remaining potential and interaction terms.

This Hamiltonian shows that the center-of-mass effectively moves on a deformed sphere with metric $h_{\mu\nu}$. While in the deeply bound limit, we find
\begin{eqnarray}
\lim_{\theta\rightarrow0}h_{\mu\nu}=2\left(
                                  \begin{array}{cc}
                                    \sin^2\beta & 0 \\
                                    0 & 1 \\
                                  \end{array}
                                \right),\label{metric of sphere}
\end{eqnarray}
which is exactly the metric tensor of a sphere. In addition, there emerges a vector potential $\mathbf{A}$ corresponding to the coupling between the center-of-mass and the relative angular momentum $L$. If we calculate the corresponding magnetic field of the vector potential $\mathbf{A}$, we obtain
\begin{eqnarray}
\mathbf{B}=\nabla\times\mathbf{A}=\frac{1}{R^2\sin\beta}\frac{\partial A_\alpha}{\partial\beta}\hat{e}_r=-\frac{\hat{e}_r}{R^2}.
\end{eqnarray}
This is exactly the magnetic field of a Dirac monopole at the center of this sphere. The relative angular momentum plays the role as monopole charge, which is quantized in a quantum theory. Thus, it is consistent with the wave function of a pair found in the quantum theory.

\textit{Physical Meaning of the Gauge Transformation.} It is natural to ask what is the meaning of gauge transformation. To answer this question, we recall that $\gamma$ is the angle between the geodesic and the local base vector $\hat{e}_\theta$. Thus the gauge transformation simply represents a local rotation of the base vector by $f(\alpha,\beta)$, and the Hamiltonian is invariant under following gauge transformation, $\gamma\rightarrow\gamma-f(\alpha,\beta)$,
$A_\alpha\rightarrow A_\alpha+\partial_\alpha f$, and $A_\beta\rightarrow A_\beta+\partial_\beta f$, where $f(\alpha,\beta)$ is an arbitrary function defined on the sphere. According to the definition of $\omega_3$, we know that $d\gamma=\omega_3 dt-A_\alpha d\alpha$. If the center of mass is carried along any a close path $\mathcal{C}$ on the sphere, the net change of $\gamma$ should not depend on the gauge choice, and it is given by
\begin{eqnarray}
\Delta\gamma=\int\omega_3dt-\oint_{\mathcal{C}}A_\alpha \mathrm{d}\alpha=\int\omega_dt-\oint_{\mathcal{C}}\mathbf{A}\cdot \mathrm{d}\boldsymbol{\ell}.
\end{eqnarray}
Besides the regular $\int\omega_3 dt$ term, the net change $\Delta\gamma$ acquires an extra geometric term related to the vector potential $\mathbf{A}$, which only depends on the path $\mathcal{C}$ and is an analog of the geometric phase introduced by Berry \cite{Berry}. Using Stokes' theorem we can write $\oint_{\mathcal{C}}\mathbf{A}\cdot \mathrm{d}\boldsymbol{\ell}=\int_\Omega\mathbf{B}\cdot\mathrm{d}\boldsymbol{s}$ which is clearly gauge invariant. That is to say, the emergent gauge field is related to this geometric effect.

\begin{figure}[tbp]\centering
\includegraphics[width=1.8in]
{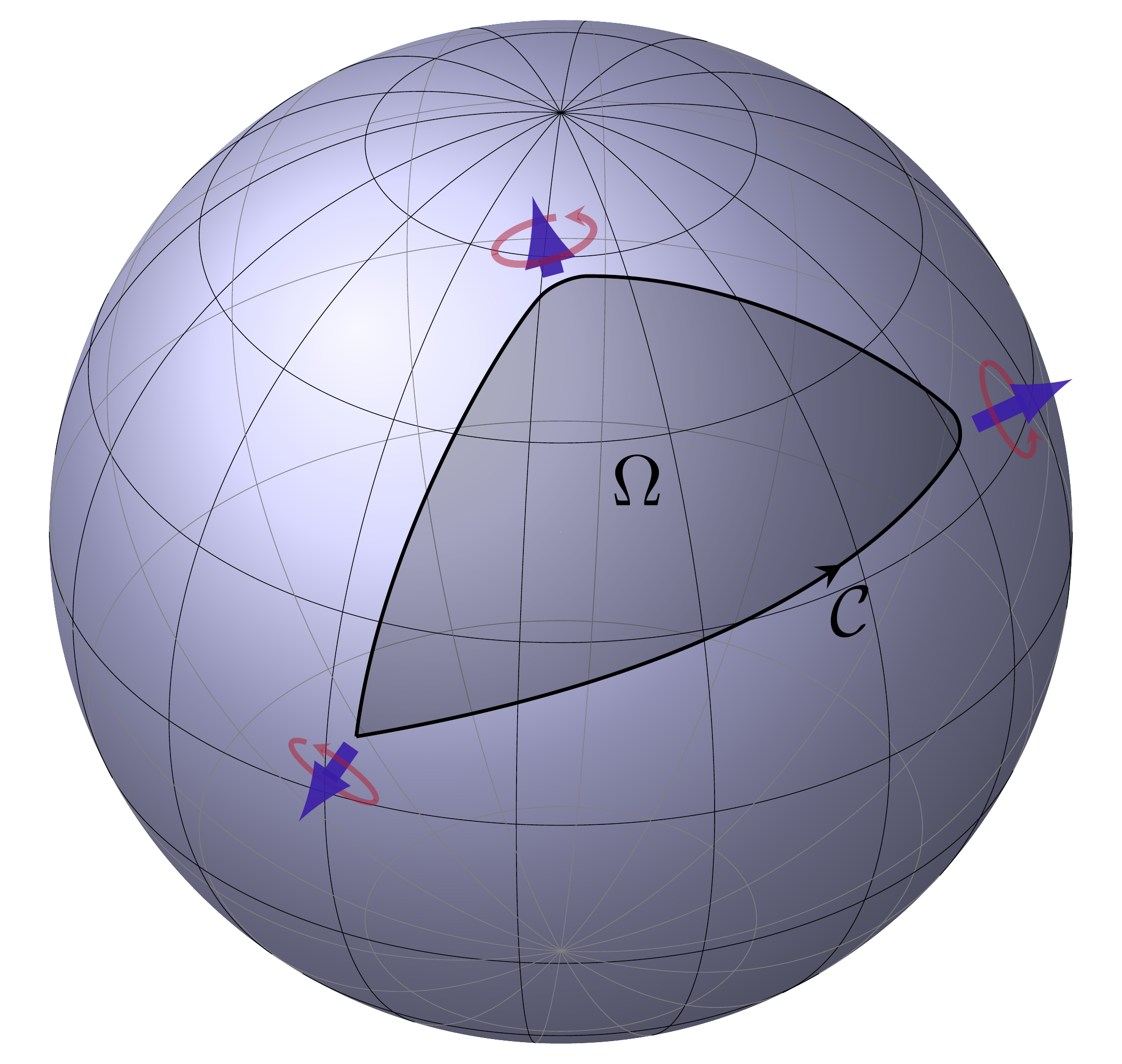}
\caption{A schematic plot of the closed path $\mathcal{C}$ and the enclosed area $\Omega$. The normal direction around which two atoms rotate varies while the center-of-mass travles along path $\mathcal{C}$.\label{fig3}}
\end{figure}

A more intuitive understanding of this gauge field is shown in Fig.\ref{fig3}. For a chiral bound state, two atoms rotate around the normal axis $\hat{n}_\text{c}$ which is always perpendicular to the local surface. Therefore, when the bound state travels along a closed loop one the surface, the direction of this normal axis varies, which gives rise to a geometric term identical to the solid angle expanded by the direction. This is similar to the Berry phase effect of a spin varying on the Bloch sphere.

\textit{Generalization to Arbitrary Manifolds.} This emergent gauge field can also be generalized to an arbitrary 2-dimensional manifold $\mathcal{M}$ with metric
\begin{eqnarray}
ds_{M}^2=g_{\mu\nu}du^\mu du^\nu=g_{11}du^1du^1+g_{22}du^2du^2.
\end{eqnarray}
For simplicity, we have chosen orthogonal coordinates $(u^1,u^2)$ such that $g_{12}=g_{21}=0$. Similar to the sphere case, we introduce coordinates $(q^1,q^2,\gamma,r)$ as shown in Fig.\ref{fig1}(b), where $(q^1,q^2)$ are the coordinates of the center of mass which is defined as the middle point of the geodesic connecting two particles, $r$ is the length of the geodesic, and $\gamma$ is the angle between the geodesic and the tangent vector $\hat{e}_1=\frac{\partial_1}{\sqrt{g_{11}}}$ at point $\mathcal{C}$.

Generally, it is difficult to write the metric tensor $ds^2$ explicitly using coordinates $(q^1,q^2,\gamma,r)$. Nevertheless, if we assume that the two particles form a deeply bound state such that we are only interested in the regime where $r$ is much smaller than any other length scales, we can expand $ds^2$ in small $r$ limit, and to the leading order, $ds^2$ becomes
\begin{eqnarray}
ds^2=\frac12{(}{r^2}d\phi^2+{dr^2}{)}+2g_{\mu\nu}dq^\mu dq^\nu
+r^2\frac{\sqrt{g}}{g_{11}}\Gamma^2_{1\mu}{\sqrt{g}}dq^\mu d\gamma.\nonumber\\\label{metric general}
\end{eqnarray}
Here $\Gamma_{\mu\nu}^\lambda$ is the Christoffel symbol of connection. Given the expression of the metric tensor, and performing similar derivation above, we find that the kinetic energy can be expressed as
\begin{align}
\hat{T}&=\frac{1}{2\sqrt{g}}(-i\partial_\mu-LA_\mu)\sqrt{g}g^{\mu\nu}(-i\partial_\nu-LA_\nu)\nonumber\\
&+\frac{L^2}{r^2}-\frac{1}{r}\frac{\partial}{\partial r}\bigg{(}r\frac{\partial}{\partial r}\bigg{)},
\end{align}
where $L=-i\partial_\gamma$ and $A_\mu=\frac{\sqrt{g}}{g_{11}}\Gamma^2_{1\mu}$. Again, we obtain a vector potential $A_\mu$ and the corresponding magnetic field is given by
\begin{eqnarray}
{\bf B}=\nabla\times\mathbf{A}=-K\hat{e}_{\perp}.
\end{eqnarray}
Here one can show that $K$ is exactly the Gaussian curvature of $\mathcal{M}$, and $\hat{e}_{\perp}=\hat{e}_1\times\hat{e}_2$ is a unit vector perpendicular to the surface. That is to say, the gauge field corresponds to a magnetic field perpendicular to the surface with the strength identical to the local Gaussian curvature. It is easy to see that the sphere case discussed above is a special example of this generalized result.

\textbf{Concluding Remarks.}
First, if one creates a surface with periodic modulated curvature, it gives rise to a magnetic flux lattice to a chiral bound state and creates topological band structure. More interestingly, considering a bound state with either angular momentum $l$ or $-l$, it can be considered as a spin-$\frac{1}{2}$ particle, and since the emergent magnetic fields are opposite for opposite angular momentums, the time-reversal symmetry is recovered which can lead to a time-reversal invariant topological insulator. Our results indicate that topological matter may also arise from nontrivial space curvature.

Secondly, our results can also be directly tested in cold atom experiments \cite{Ho}. For instance, considering a Rydberg atom whose outermost shell electron is excited to a highly excited state with wave function $\varphi_\text{e}({\bf r})$, the interaction between other ground state atoms and the Rydberg atom is proportional to the density of the excited electron $|\varphi_\text{e}({\bf r})|^2$ \cite{Rydberg1,Rydberg2}, where the most attractive part is a thin shell of sphere centered at the position of the ion. It has been observed that a few atoms can be trapped by this potential shell \cite{Rydberg1,Rydberg2}. These trapped atoms basically live on a two-dimensional sphere and interact via short-range potential, as our model requires. This will be an ideal system to test our results.

{\it Acknowledgements.} We thank Ran Qi, Zhenhua Yu and Pengfei Zhang for helpful discussions. This work is supported by Tsinghua University Initiative Scientific Research Program, and NSFC Grant No. 11174176(HZ), No. 11325418 (HZ).

\begin{widetext}

\vspace{0.05in}

\section{appendix}
The matrix elements of metric tensor $g_{ij}$ and $h_{\mu\nu}$ are
\begin{eqnarray}
g_{ij}=
2\left(
  \begin{array}{cccc}
    \sin^2\beta(\cos^2\gamma+\cos^2\frac\theta2\sin^2\gamma)+\sin^2\frac\theta2\cos^2\beta & -\sin^2\frac{\theta}{2}\sin\beta\sin\gamma\cos\gamma & \sin^2\frac{\theta}{2}\cos\beta & 0 \\
    -\sin^2\frac{\theta}{2}\sin\beta\sin\gamma\cos\gamma & \sin^2\gamma+\cos^2\frac\theta2\cos^2\gamma & 0 & 0 \\
    \sin^2\frac{\theta}{2}\cos\beta & 0 & \sin^2\frac{\theta}{2} & 0 \\
    0 & 0 & 0 & \frac14 \\
  \end{array}
\right),\label{metric_g}\nonumber
\end{eqnarray}
\begin{eqnarray}
h_{\mu\nu}=2\left(
              \begin{array}{cc}
                \sin^2\beta(\cos^2\gamma+\cos^2\frac\theta2\sin^2\gamma) & -\sin^2\frac\theta2\sin\beta\sin\gamma\cos\gamma \\
                -\sin^2\frac\theta2\sin\beta\sin\gamma\cos\gamma & \sin^2\gamma+\cos^2\frac\theta2\cos^2\gamma \\
              \end{array}
            \right).\label{metric_h}\nonumber
\end{eqnarray}

The explicit expressions for three angular momentums $J_i$ in terms of Euler angles are
\begin{eqnarray}
J_1&=&i\bigg{(}\frac{\cos\gamma}{\sin\beta}\frac{\partial}{\partial\alpha}-\sin\gamma\frac{\partial}{\partial\beta}-\cot\beta\cos\gamma\frac{\partial}{\partial\gamma}\bigg{)}\nonumber\\
J_2&=&i\bigg{(}-\frac{\sin\gamma}{\sin\beta}\frac{\partial}{\partial\alpha}-\cos\gamma\frac{\partial}{\partial\beta}+\cot\beta\sin\gamma\frac{\partial}{\partial\gamma}\bigg{)}\nonumber\\
J_3&=&-i\frac{\partial}{\partial\gamma}.\nonumber
\end{eqnarray}

The expressions for $A$, $B$ and $C$ are
\begin{eqnarray}
A=\frac{1}{8}\bigg{(}\frac{1}{\cos^2\theta/2}+1\bigg{)},\quad
B=\frac{1}{8}\bigg{(}\frac{2}{\sin^2\theta/2}\frac{1}{\cos^2\theta/2}-1\bigg{)},\quad
C=\frac{\tan^2\theta/2}{16}.\nonumber
\end{eqnarray}
\end{widetext}


\begin{thebibliography}{99}
\bibitem{Euler angles}
Here we will use the so-called $z-y-z$ convention to define Euler angles, which is commonly used in quantum mechanics.
\end{thebibliography}
\end{document}